# "Name that manufacturer". Relating image acquisition bias with task complexity when training deep learning models: experiments on head CT

Giorgio Pietro Biondetti [1]
Romane Gauriau, PhD [1]
Christopher P. Bridge, PhD [1]
Charles Lu [1]
Katherine P. Andriole KP, PhD [1,2]

[1]MGH & BWH Center for Clinical Data Science, Boston, MA, USA
[2]Department of Radiology, Brigham and Women's Hospital, Harvard Medical School, Boston, MA, USA
gbiondetti@mgh.harvard.edu

## Abstract

As interest in applying machine learning techniques for medical images continues to grow at a rapid pace, models are starting to be developed and deployed for clinical applications. In the clinical AI model development lifecycle (described by Lu *et al*. [1]), a crucial phase for machine learning scientists and clinicians is the proper design and collection of the data cohort. The ability to recognize various forms of biases and distribution shifts in the dataset is critical at this step. While it remains difficult to account for all potential sources of bias, techniques can be developed to identify specific types of bias in order to mitigate their impact. In this work we analyze how the distribution of scanner manufacturers in a dataset can contribute to the overall bias of deep learning models. We evaluate convolutional neural networks (CNN) for both classification and segmentation tasks, specifically two state-of-the-art models: ResNet [2] for classification and U-Net [3] for segmentation. We demonstrate that CNNs can learn to distinguish the imaging scanner manufacturer and that this bias can substantially impact model performance for both classification and segmentation tasks. By creating an original synthesis dataset of brain data mimicking the presence of more or less subtle lesions we also show that this bias is related to the difficulty of the task. Recognition of such bias is critical to develop robust, generalizable models that will be crucial for clinical applications in real-world data distributions.

# 1. Introduction

The development of AI algorithms using supervised learning requires large and heterogeneous datasets. Several important considerations should be taken into account when creating medical imaging datasets for machine learning (ML) [4].

A cohort has to be selected while considering multiple factors that could introduce bias in training the ML model. Such factors could include ensuring sufficient diversity and minimal imbalance in patient demographics (such as age, gender, ethnicity, etc.), exclusion criteria (edge cases, possible confounders), and characteristics of the data acquisition process (manufacturer, model of the machine, software version, acquisition protocols, etc.), particularly when data comes from a single institution where the distribution of patients and machines may be limited.

In this work we study how the scanners used in image acquisition can impact the performance and generalizability of deep learning models. We first show that a shallow neural network can easily learn to discriminate between images acquired on scanners from two different manufacturers, which leads us to hypothesize that scanners can possess different "signatures", which manifest as subtle differences in image noise and/or patterns. This variability could be introduced at different stages or configurations in the image creation process such as scanner protocols/settings, image reconstruction algorithms, or image processing operations. However, it is not yet clear how these signatures could affect further downstream image analysis or model training; in particular, the impact on the generalizability of neural network models for different types of computer vision tasks.

To test our hypothesis, we design an experimental setup with different classification and segmentation tasks of different levels of difficulty to be learned by different types of neural network architectures. For the classification task, we use a dataset of non-contrast CT (NCCT) brain images in which we introduce spheres of different sizes and contrast, thus mimicking brain tumors. By controlling for size and contrast, we are able to control the difficulty of learning to detect them. Thus, we can define easy, medium, and difficult/hard tasks. We use the same dataset for a segmentation task with the goal being to segment the spheres of increasing detection difficulties.

We use data collected retrospectively from two institutions within the Mass General Brigham ecosystem: Massachusetts General Hospital (MGH) and Brigham and Women's Hospital (BWH). Our setting is similar to other clinical environments where AI algorithms are developed and deployed, in which multi-institutional data is difficult to access due to regulatory and privacy issues [4].

As our experiments show, differing scanner manufacturers can impart bias through characteristics of image acquisition that can impact ML model performance in simple classification and segmentation tasks. Detecting this form of bias is important so that it may be rectified or mitigated.

## 2. Prior Work

Large datasets are crucial for modern deep learning techniques and have enabled considerable progress in the field, in particular for imaging research. When building a dataset, the general aim is to attempt to encapsulate the (potentially infinite) combinations of variables that comprise the true, underlying data distribution into a finite set of samples for training [4]. In an attempt to satisfy these requisites, many datasets have been created by research groups.

Several previous works have investigated dataset bias in the context of computer vision. Torralba and Efros showed that every dataset has its own unique "signature" [5]. They showed with their simple experiment, termed "Name the Dataset", how training a classifier on a random sample of 1000 images from 12 different datasets can achieve a classification accuracy of 39% (random chance being only 1/12 = 8%). Following in the same vein, Ashtraf *et al*. [6] trained a "Name the Study" classifier using four brain MRI datasets (ADNI 1, ADNI 2, ADNI GO [7], and AIBL [8]) achieving an overall classification accuracy of 66% (random chance being 1/4 = 25%). Another study quantified dataset bias using 15 neuroimaging large datasets, which were then used for the selection of a training set for autism prediction [9]. Gender differences were evaluated in chest X-Ray images [10], showing that models trained on balanced datasets have overall better performances than models trained on only one gender.

In this paper, starting with a similar approach as in the aforementioned works, we address how the scanner manufacturer can play a central role in biasing deep learning models. We want to show that it is important to take into account all possible sources of bias and that one of the possible solutions to avoid them is balancing the distribution in the dataset.

## 3. Methods

### 3.1. Data Source

This research was approved by our Institutional Review Board (IRB) prior to acquiring any patient data. Patient consent was waived by IRB. Our dataset consisted of 782 non-contrast CT (NCCT) head studies: 50% acquired on General Electric (GE) scanners and 50% on Siemens scanners (different models were included). All images were collected from two institutions: Massachusetts General Hospital (MGH) and Brigham and Woman's Hospital (BWH). Exams were carefully selected to maintain similar distributions in age and gender across both manufacturers, using information from the respective Digital Imaging and Communications in Medicine (DICOM) metadata. The axial series with 5mm slice thickness was selected using the Series Description DICOM tag. For the examinations lacking a 5mm axial series, another axial series was selected with <5mm slice thickness (1.25mm – 2.5mm) and resampled using linear interpolation to obtain a series with 5mm slice thickness. Each slice of every exam was resampled to 256 x 256 using linear interpolation to decrease training time. All models were trained using P100 Nvidia GPUs and all data saved in NIFTI format to allow faster processing speed during training.

## 3.2. Synthetic Dataset Generation

Our data cohort, detailed in Table 1, was split into training (70%), validation (9%) and test (21%) sets while maintaining balance across manufacturers and class labels: positives (image with synthetic sphere/'tumor') and negatives (without a synthetic sphere).

| Division | Total | Manufacturers | "positives" | "negatives" |
|---|---|---|---|---|
| Training | 552 (70%) | 276 GE | 138 | 138 |
| | | 276 Siemens | 138 | 138 |
| Validation | 68 (9%) | 34 GE | 17 | 17 |
| | | 34 Siemens | 17 | 17 |
| Testing | 162 (21%) | 81 GE | 40 | 41 |
| | | 81 Siemens | 40 | 41 |
| TOTAL | 782 | 391 GE | 195 | 196 |
| | | 391 Siemens | 195 | 196 |

Table 1. NCCT dataset divided into train-validation-test sets. Positives and negatives columns indicate images with and without spheres, respectively.

A synthetic dataset was created for each experiment described below:

- *Original*: this dataset is composed of 782 unaltered NCCTs labeled with their respective scanner manufacturer. For each examination, 35 slices have been selected (verifying that the brain was inside this region of interest (ROI)) and resampled to give a volume with dimensionality of 256 x 256 x 35 as input to the ML models.

- *Easy_Spheres*: this dataset is a modification of the *Original* dataset. To create this dataset, we introduced a single sphere inside the brain of half of the examinations, maintaining balance with respect to manufacturer, so that half of the spheres were introduced into GE studies and half into Siemens. Spheres varied with a radius between 21-26mm and pixel intensity was drawn from a normal distribution with mean 50 Houndsfield Unit (HU) and standard deviation 2 HU, simulating a mass or bleed in the patient brain (example in Figure 1, top). Those exams with spheres were labeled *positive* and those without labeled *negative*. The spheres were placed randomly inside the brain volume using an iterative process: first the skull was removed, then a point in the brain was randomly selected and a sphere created using the point coordinates; if the sphere was entirely inside the brain volume the sphere was placed, if not a new center was selected. The center locations were finally saved to allow the radius parameter to be varied for subsequent datasets.

- *Medium_Spheres*: for the creation of this dataset, we initially started from the *Easy_Spheres* dataset, and we maintained the same radius, mean and stdev of spheres for 80% of the *positive* series. For the remaining 20%, we varied the radius between 13-17mm and drew pixel intensities from a Normal distribution with mean 40 HU and standard deviation 2 HU (example in Fig. 1, bottom). The smaller spheres and decreased contrast make the detection task slightly more difficult.

- *Hard_Spheres*: the same procedure as in the *Medium_Spheres* was used for the creation of this dataset but splitting the *positive* cases in which 65% maintain the same radius but placing smaller spheres for the remaining 35%.

- *Skull*: starting from the raw data, we automatically removed the skull from each series using an in-house skull-stripping algorithm and saved it as a binary brain mask (Fig. 2). Each series-mask pair was then resampled along the craniocaudal axis to reduce the number of slices from 35 to 32. This was done to decrease GPU memory usage and the number of parameters needed in each model.

- The '*Spheres*' and '*Skull*' datasets described above were used during the experiments described in Section 3.3. We add the labels _GE or _Si to each of the datasets to represent the sets of imaging studies coming from a GE or a Siemens scanner respectively.

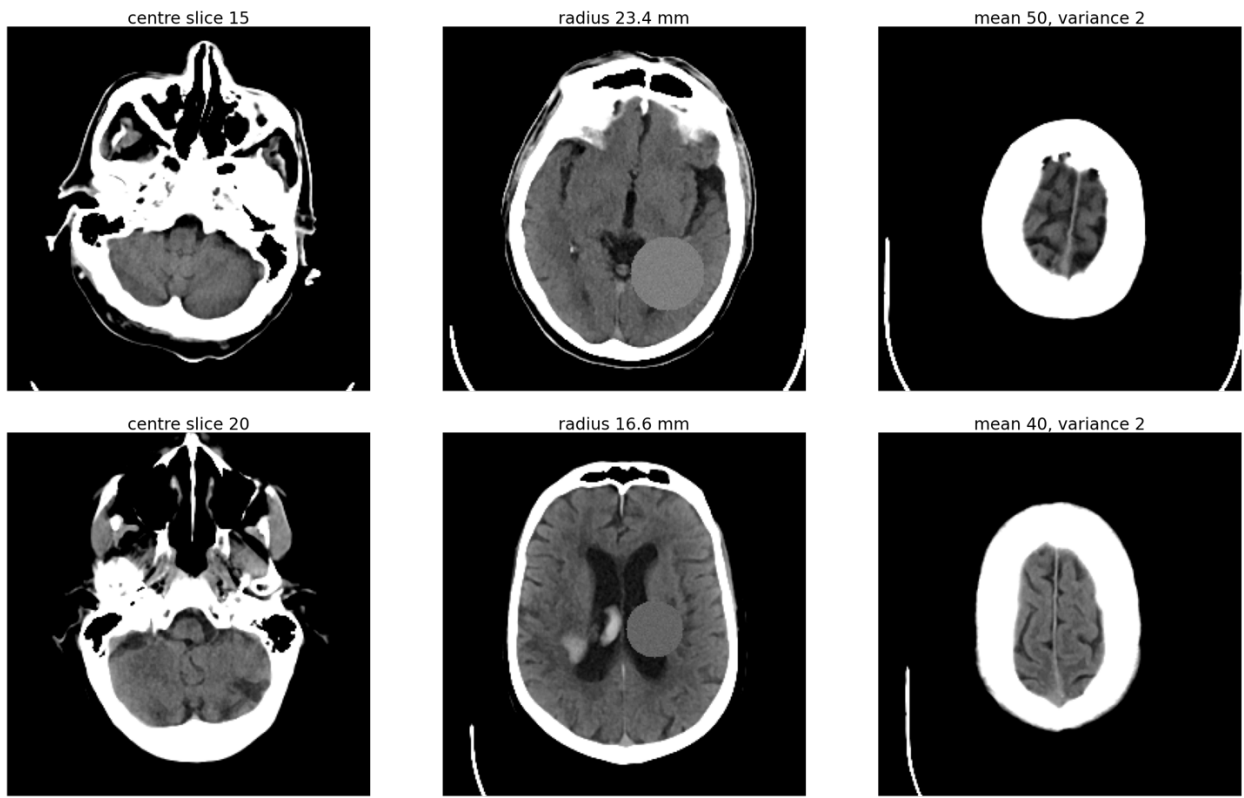

**Fig. 1**: Top: Sample of positive cases of *Easy_Sphere*. Bottom: sample of the smaller sphere positive cases of both *Medium_Sphere* and *Hard_Sphere* datasets.

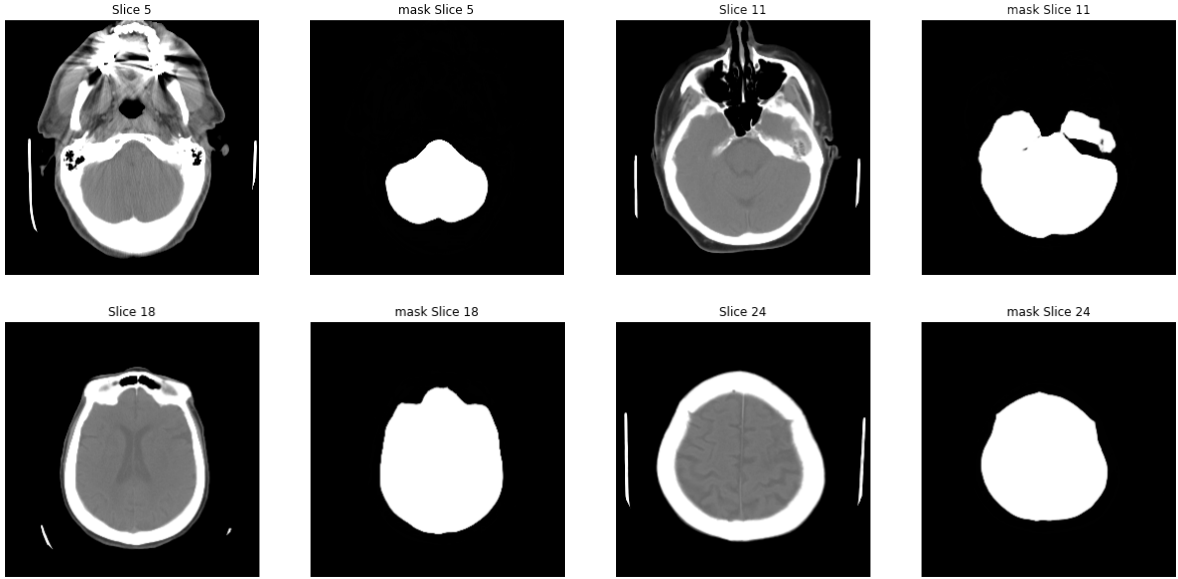

**Fig. 2**: Examples of NCCT slice and corresponding brain mask for the Skull dataset.

### 3.3. Experiments

In Section 3.3.1 we test whether we can classify an image as coming from either a GE or a Siemens scanner using the *Original* dataset to support the existence of manufacturer bias. We then analyze a specific classification task using a dataset in which we have introduced perturbations and can control them. We synthesize three similar tasks with different degrees of "difficulty" using the datasets: *Easy_Sphere, Medium_Sphere, Hard_Sphere* described above. This experiment (Section 3.3.2) examines whether the effects of the manufacturer bias are correlated with the difficulty of the task itself. In Section 3.3.3 we analyze a segmentation task using an approach similar to Section 3.3.2, in which we utilize the *Skull* dataset.

#### 3.3.1. "Name the Manufacturer"

For this experiment we trained a shallow CNN to classify a study according to the manufacturer of the scanner from which it was acquired. The input to the model was a random selection of three slices from each series of the *Original* dataset, which were stacked together randomly. This random stacking was performed to ensure each stack was independent from each other as well as from any anatomical ordering that could create bias in the model input. As we were interested in overall image patterns rather than specific anatomical features we preprocessed our images using a broad window-level. In particular, the pixel intensities (rescaled to the Hounsfield Unit) were mapped from a window/level of 40/400 respectively to the interval [-1, 1], and intensities outside were clipped (examples in Fig. 3).

The deep learning model used for this task consisted of three CNN blocks, each composed of one layer of 3x3 convolution followed by batch normalization and max pooling layers with a 2x2 factor. The number of convolution filters was (32, 64, 128) for each block, respectively. The last layer was a dense layer with a sigmoid activation function. The model was trained using cross-entropy loss and the Adam [11] optimizer with a learning rate of $5 \times 10^{-3}$ and a batch size of 32.

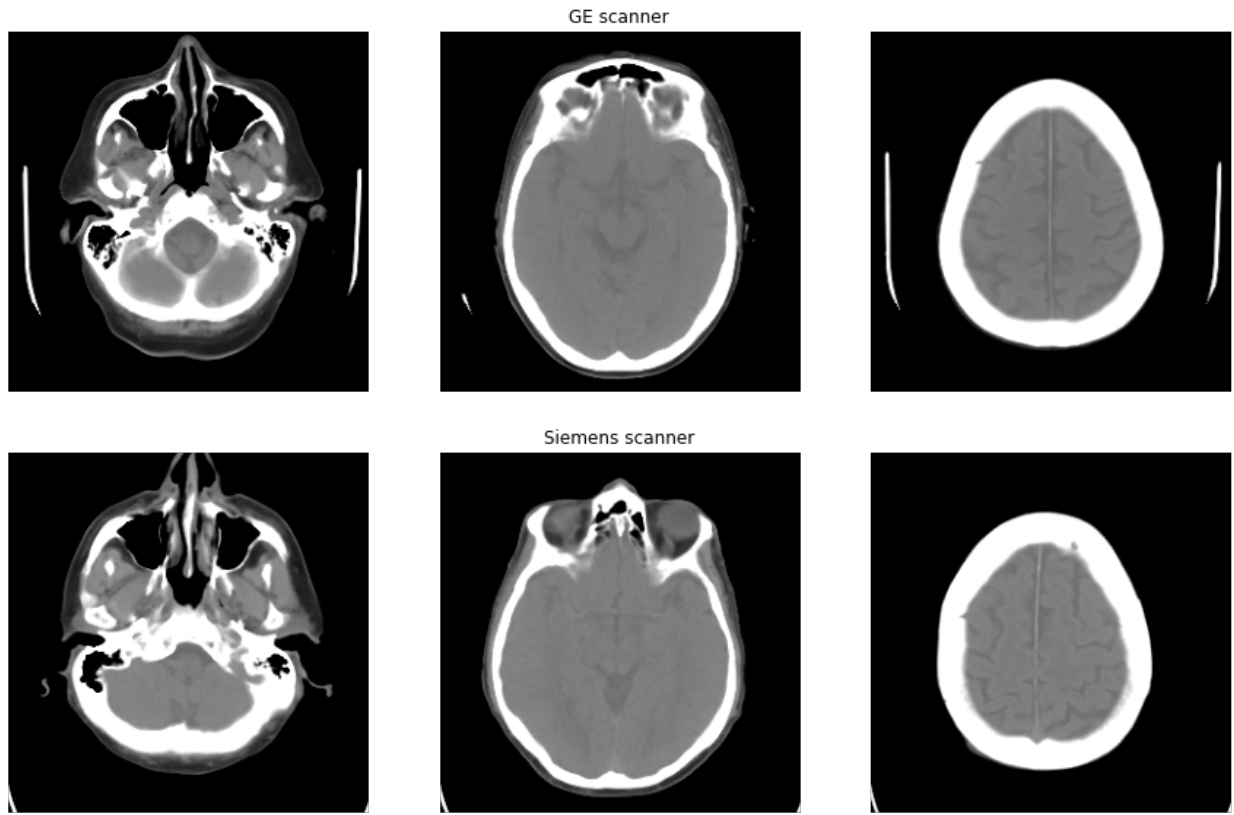

**Fig. 3**: Input samples from the *Original* dataset. Each row shows different slices of a same input series.

### 3.3.2. Sphere classification

**Model Architecture**
For the sphere/tumor classification experiments, we used a 3D variant of the ResNet model architecture [12]. In this architecture, the first convolutional block had a filter of dimension (7, 7, 7) with a stride of two followed by max pooling of dimension (2, 2, 2), then followed by four scale steps with (32, 64, 128, 256) channels each. Each scale is composed of two residual blocks and each block is composed of two convolutional layers with filter of dimension (3, 3, 3) with a down-sample stride of 2 and a max pooling layer of dimension (2, 2, 2). The last part of the network is composed of an average pooling layer followed by a dense layer with 1000 neuron units and a sigmoid activation function. This model was trained using cross-entropy loss, an Adam optimizer with learning rate $5 \times 10^{-5}$ and batch size of 32.

The model takes as input the NCCT volume of dimension (256, 256, 35). The HU intensities of the input pixels were mapped from a window and level of 100 and 50 to [0, 1] to enhance the brain window and to mimic the contrast used for brain examination.

**Experimental Settings**
Using the datasets with increasing levels of difficulty, we evaluated the effects of the manufacturer on the classification task. Using *Easy_Spheres* we trained 10 models and then selected the best six models based on accuracy and the area under the curve (AUC) of the receiver operating curve (ROC) for the validation set. We then split this dataset into *Easy_Spheres_GE* and *Easy_Spheres_Si* maintaining the same distribution of positive (with spheres) and negative cases. For each manufacturer-specific dataset we then trained 10 models and selected the best six models using the same criteria as before. We applied the same approach using *Medium_Spheres* and *Hard_Spheres*. The only differences that we encountered were that as the task became more difficult, fewer models converged so we had to train much more models to obtain a group of 6 converged models. Models trained only on Siemens series were converging half the time as compared to models trained only on GE series. A total of about 300 models were trained to obtain the results described in Section 4.

### 3.3.3. Segmentation

**Model Architecture**

A 3D variant of the state-of-the-art U-Net architecture was used to quantify bias in a segmentation task using the *Skull* dataset. The model architecture consisted of five compression stages with (8, 16, 32, 64, 128) channels respectively, followed by the decompression stages that mirror them. Each stage is formed by two convolutional layers with kernel size of 3 in each dimension, followed by a pooling layer of dimension 2. The final layer uses a sigmoid activation function and outputs the predicted mask corresponding to the input series. The network was trained using the Dice loss [13] and Adam optimizer with learning rate of $1 \times 10^{-4}$. The model takes as input the NCCT volume of dimension (256, 256, 32). The same normalization process as in the classification experiments was used.

**Experimental Settings**

After training 5 models with *Skull* for hyperparameter optimization, we observed that high Dice scores (above 0.96) were consistently easy to achieve. We then trained on *Skull_GE* and *Skull_Si*, using the same approach implemented to analyze the classification task. A total of 20 models were trained to obtain the results described in Section 4.

## 4. Results and Discussion

### 4.1. "Name the Manufacturer!"

This first experiment shows how easy the task of learning image acquisition parameters including the manufacturer is for even a shallow neural network. Results in Table 2 show the average performance of two ensembles of models composed of five models each. It is interesting to note that in both groups not a single image coming from a Siemens machine was misclassified.

| Best 2 models | Test accuracy | AUROC | Siemens | GE |
|---|---|---|---|---|
| Group 1 | 98.11 ± 0.65% | 0.99 ± 0.01 | 100% | 96.21 ± 1.31% |
| Group 2 | 98.86 ± 1.25% | 0.99 ± 0.01 | 100% | 97.72 ± 2.21% |

Table 2. Results for the two groups of models trained on the manufacturer label.

## 4.2. Sphere Classification

We selected the best six models tested for each task and subtask. Model selection was based upon assessing the behavior of the training and validation loss and accuracy curves for overfitting, evaluation accuracy, sensitivity/specificity, and area under the receiver operating curve (AUROC) performance metrics and evaluating the convergence of the models. The final provided metrics are averaged over the six selected models.

Table 3 shows the results for the experiment using the *Spheres* datasets that included a mixture of both manufacturers for training. Comparing the "Test" column (mixed manufacturer) with the "Test on GE" and "Test on Siemens" columns, it can be noted that overall models have slightly better performances on GE series when trained on both manufacturers.

Table 4 and Table 5 show the other two experiments in which we trained the same models using *Spheres_GE* and *Spheres_Si* during training, and then tested both models on both manufacturers. datasets. Some interesting observations can be made. Models trained on *Spheres_GE* perform better on the GE test set; for both medium and hard tasks, they perform better than when trained on the whole mixed manufacturer dataset. When testing on Siemens studies, they have very poor specificity compared to sensitivity. For the models trained on *Spheres_Si*, all models have lower performances compared with the models trained on *Spheres_GE*. On the easy task, the performances on both Siemens and GE series do not differ significantly. Unexpectedly, on the medium task the model performs better on the *Spheres_GE* test set. Overall on both test sets, the models have worse sensitivity compared to the specificity.

| Metrics | Validation - Mixed | Test - Mixed | Test on GE | Test on Siemens |
|---|---|---|---|---|
| Easy task | | | | |
| Accuracy | 98.8 ± 0.0 | 99.0 ± 0.0 | 99.6 ± 0.0 | 98.4 ± 0.0 |
| Specificity | 98.5 ± 0.0 | 98.1 ± 0.0 | 99.6 ± 0.0 | 96.6 ± 0.1 |
| Sensitivity | 99.0 ± 0.0 | 99.8 ± 0.0 | 99.6 ± 0.0 | 100 ± 0.0 |
| AUROC | 99.9 ± 0.0 | 99.6 ± 0.0 | 100 ± 0.0 | 99.2 ± 0.0 |
| Medium task | | | | |
| Accuracy | 85.3 ± 0.1 | 84.1 ± 0.1 | 86.6 ± 0.1 | 81.5 ± 0.1 |
| Specificity | 89.7 ± 0.9 | 88.1 ± 0.6 | 90.8 ± 0.5 | 85.4 ± 0.8 |
| Sensitivity | 80.9 ± 0.4 | 80.8 ± 0.2 | 82.5 ± 0.1 | 77.6 ± 0.3 |
| AUROC | 90.0 ± 0.0 | 88.7 ± 0.0 | 91.4 ± 0.0 | 85.8 ± 0.1 |
| Hard task | | | | |
| Accuracy | 70.6 ± 1.8 | 68.2 ± 1.5 | 69.1 ± 1.8 | 67.1 ± 1.3 |
| Specificity | 77.4 ± 10.7 | 76.5 ± 8.9 | 77.5 ± 8.4 | 75.4 ± 9.6 |
| Sensitivity | 73.7 ± 9.9 | 60.2 ± 8.2 | 61.0 ± 7.9 | 59.4 ± 8.6 |
| AUROC | 76.3 ± 1.8 | 77.9 ± 1.5 | 80.1 ± 1.7 | 75.8 ± 1.3 |

Table 3. Average performance metrics for six models trained on mixed manufacturer *Spheres* datasets, tested on mixed, GE only and Siemens only cases.

| Metrics | Validation GE | Test GE | Test Siemens |
|---|---|---|---|
| Easy task | | | |
| Accuracy | 99.5 ± 0.0 | 98.6 ± 0.0 | 83.3 ± 0.5 |
| Specificity | 100 ± 0.0 | 97.9 ± 0.0 | 66.3 ± 2.0 |
| Sensitivity | 99.1 ± 0.0 | 99.2 ± 0.0 | 100 ± 0.0 |
| AUROC | 99.8 ± 0.0 | 98.7 ± 0.0 | 91.9 ± 0.1 |
| Medium task | | | |
| Accuracy | 95.7 ± 0.1 | 92.5 ± 0.1 | 70.0 ± 0.7 |
| Specificity | 98.0 ± 0.0 | 92.1 ± 0.4 | 47.9 ± 3.4 |
| Sensitivity | 93.5 ± 0.2 | 93.1 ± 0.1 | 91.5 ± 1.5 |
| AUROC | 98.2 ± 0.0 | 97.4 ± 0.0 | 82.6 ± 1.0 |
| Hard task | | | |
| Accuracy | 92.9 ± 0.3 | 88.1 ± 1.0 | 67.9 ± 0.5 |
| Specificity | 91.2 ± 1.2 | 83.3 ± 3.0 | 45.8 ± 6.9 |
| Sensitivity | 94.4 ± 0.0 | 92.7 ± 0.4 | 89.4 ± 4.2 |
| AUROC | 96.1 ± 0.0 | 95.7 ± 0.2 | 78.4 ± 0.7 |

Table 4. Average performance metrics for models trained on *Spheres_GE*.

| Metrics | Validation Siemens | Test GE | Test Siemens |
|---|---|---|---|
| Easy task | | | |
| Accuracy | 97.5 ± 0.2 | 95.7 ± 0.7 | 95.9 ± 0.6 |
| Specificity | 97.1 ± 0.2 | 98.3 ± 0.1 | 97.1 ± 0.2 |
| Sensitivity | 97.9 ± 0.2 | 93.1 ± 1.5 | 94.7 ± 1.2 |
| AUROC | 99.1 ± 0.0 | 97.5 ± 0.3 | 96.9 ± 0.3 |
| Medium task | | | |
| Accuracy | 84.3 ± 0.5 | 84.0 ± 0.2 | 78.0 ± 0.2 |
| Specificity | 88.2 ± 2.6 | 95.0 ± 0.5 | 88.3 ± 1.0 |
| Sensitivity | 80.2 ± 1.2 | 73.2 ± 0.8 | 67.9 ± 0.8 |
| AUROC | 90.1 ± 0.3 | 89.0 ± 0.1 | 84.4 ± 0.5 |
| Hard task | | | |
| Accuracy | 65.6 ± 0.5 | 58.2 ± 0.6 | 65.2 ± 0.4 |
| Specificity | 73.5 ± 3.8 | 90.0 ± 2.8 | 88.3 ± 0.9 |
| Sensitivity | 57.3 ± 1.4 | 27.2 ± 6.9 | 42.7 ± 2.9 |
| AUROC | 73.1 ± 0.5 | 71.7 ± 1.0 | 77.2 ± 0.3 |

Table 5. Average performance metrics for models trained on *Spheres_Si*.

In Fig. 4 we show the Receiver Operating Characteristic (ROC) curves, obtained by averaging the outputs from all models, for each series grouped by task and experiment. GE trained models appear to perform better on the GE test set, and on the Siemens test set, there is no significant difference from models trained on both manufacturers. It is interesting that even models trained on Siemens perform better on GE than on Siemens in the medium task.

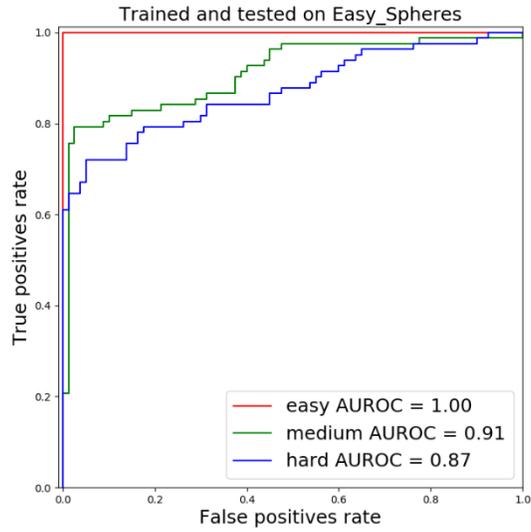

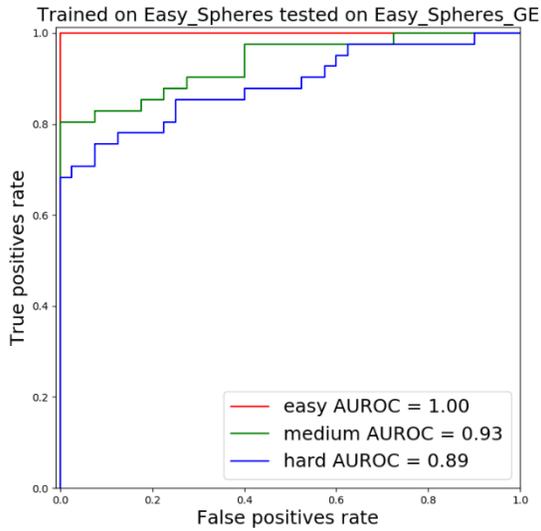
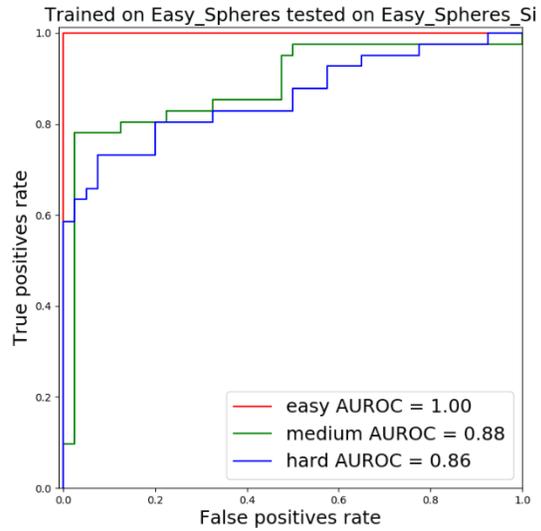

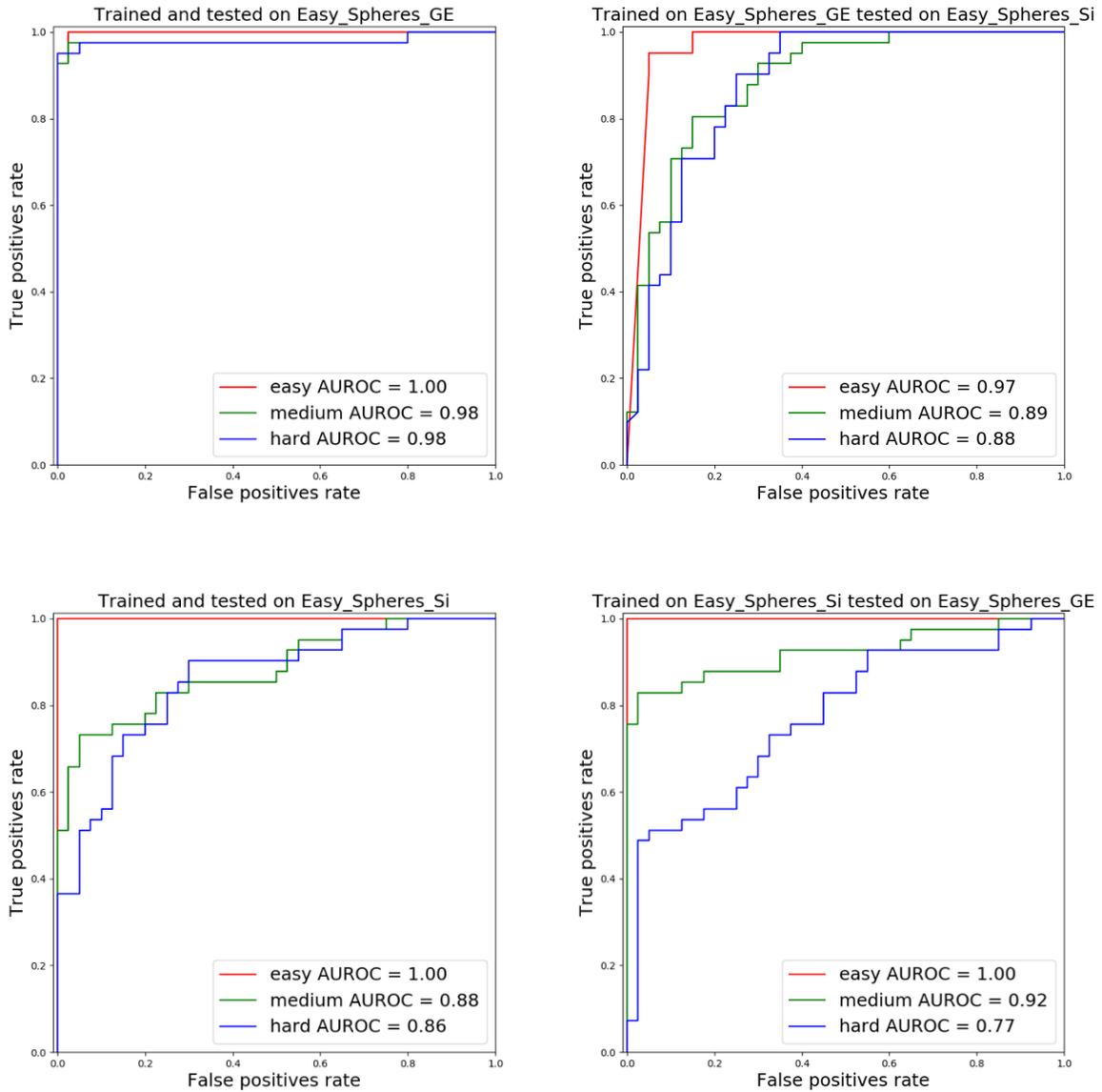

**Fig. 4**: ROC curves for averaged model outputs for each task and different manufacturer.

### 4.3. Segmentation

Table 6 shows the average Dice coefficient scores for each group of models. We observe that models trained on *Skull* (includes mixed manufacturers) perform better on the test set than models trained on only a single manufacturer. We also note that this task can be denoted as "easy" as all models converge to approximately the same performance.

| | Models Trained on GE+Siemens | | | |
|---|---|---|---|---|
| | Full Val Set | GE Test Set | Siemens Test Set | Full Test Set |
| Dice | 0.9848 | 0.9859 | 0.9704 | 0.9784 |
| | Models Trained on GE | | | |
| | GE Val Set | GE Test Set | Siemens Test Set | Full Test Set |
| Dice | 0.9712 | 0.9689 | 0.9664 | - |
| | Models Trained on Siemens | | | |
| | Siemens Val Set | GE Test Set | Siemens Test Set | Full Test Set |
| Dice | 0.9702 | 0.9564 | 0.9715 | - |

Table 6. Average Dice scores for models trained respectively with *Skull, Skull_GE* and *Skull_Si.*

# 5. Conclusions

In this paper we demonstrate the influence of manufacturer distribution in ML datasets and analyze how this imbalance can affect classification and segmentation tasks. First, we detect this bias, showing how easy it is for the model to learn the manufacturer imaging parameters. We show that manufacturer distribution can affect binary classification using a detection task with increasing levels of difficulty. We also evaluate a simple segmentation task, showing that manufacturer distribution can affect model performance, although to a lesser extent than in the classification task. We observe that models trained on a dataset consisting of a balanced distribution of scanner manufacturers perform better than models trained only on a single manufacturer.

We emphasize the importance of taking into account the different sources of bias when developing deep learning models that are to be used in a clinical setting, as their performance can vary depending on multiple factors such as the acquisition process and parameters. As our experiments showed, this is particularly true for tasks that are difficult to learn (subtle intensity contrasts, models hardly converging). It has to be noted that this study has some limitations. The dataset we use keeps relatively small and we focused on one single modality and one single anatomy. Moreover, our dataset contained images from only two manufacturers with few different scanner models. Further analysis could be made using larger datasets, different imaging techniques and different body structures. It would also be interesting to see whether if there are dataset shifts between data acquired by the same scanner (manufacturer and model) in different institutions.

While more generalizable approaches are being developed (like federated learning [14][15][16] and generation of synthetic data with GANs [17][18]) and data homogenization techniques are being explored [19][20][21], it is important to recognize that certain biases will exist in every data cohort (like age and gender [22][23]). Standardization of image acquisition could also help in developing and building more robust and generalizable AI tools. This standardization would require that manufacturers agree on the specifications of their acquisition process, narrowing down the differences between machines produced by competitors on the market. One other solution could be the standardization of homogenization techniques and data preprocessing, that would definitely lead to a more effective generalization of AI models.